# Vector network analysis based on wideband direct photonic digitizing

Zhengtao Jin, Min Ding, Jianping Chen, and Guiling Wu*

Abstract—Vector network analyzers (VNAs) have become one of the indispensable tools in various fields, such as medicine, material, geology, communication, and etc, due to the capacity of measuring and analyzing the response of the object under test. Conventional VNAs, commonly based on mixing architecture, have to make compromises among accuracy, dynamic range, and bandwidth. In this paper, we propose a wideband photonic vector network analyzer (PVNA) based on wideband direct photonic digitizing. Ultrastable optical trains directly undersample the response signals from objects under test, followed by electro-optic conversion and quantization, obviating the intricate down-conversion procedures in traditional VNAs. Adopting existing commercial devices, the proposed PVNA can not only extend the measurement frequency range to 110 GHz or higher but also achieve a high linearity and accuracy performance. To validate the theoretical analysis, we establish an experimental PVNA, realizing a measurable frequency span of up to 40 GHz and a dynamic range of more than 120 dB. A measured scattering parameters of a bandpass filter The experimental result is well consistent with that of a commercial network analyzer.

State Key Laboratory of Advanced Optical Communication Systems and Networks, Department of Electronic Engineering, Shanghai Jiao Tong University, Shanghai 200240, China.

## I. INTRODUCTION

Maxwell's equations indicate that when an object under test (OUT) is stimulated by electromagnetic waves, the response electromagnetic waves carry with wealthy information representing their inherent characteristics [1], [2]. Thanks to this theory, engineers develop chips, devices, and systems of RF & microwave, terahertz, or optics [3]–[8]; medical staffs address the detection of human cancer cells, the tomographic scanning, as well as the perception for health monitoring and care [9]–[12]; scientists achieve materials charact erization, geological exploration, sensing & imaging, etc [13]–[19]. Behind not only these applications but also other plenty of scientific and engineering scenarios, vector network analyzers (VNAs) have become one of the indispensable tools that record and analyze the incident and response electromagnetic signals [9], [20]–[24]. However, with boosting the pace of social development, these emerging new science and technology also put forward urgent and stringent measurement requirements in terms of accuracy, bandwidth, and dynamic range. For example, for the sake of advances in material science and engineering, various metamaterials, such as graphene and liquid crystal polymer, have stepped into W-band (75 - 110 GHz) [25]–[27]. To drive the deployment of 5G and the development of 6G communication systems, the basic components, like amplifiers, filters, etc, have to shift the working frequency band to 50GHz or higher [28]–[31]. The autonomous vehicles bring us an unprecedented driving experience, and its core technology, the automotive radar system, needs to perform at the 77 GHz frequency band to achieve high-precision positioning [32], [33]. On the other hand, a large measurable dynamic range can cover more application scenarios. For instance, tomographic as well as sensing systems require VNAs with a dynamic range of 70 dB, since the strong reflection occurring at the tissue-medium boundary and the attenuation by the lossy tissues leave a weak response signal [10]. From materials, devices to systems, the evaluation, testing, and optimization of the advanced fields above are essentially dependent on VNAs with abilities of accuracy, wide bandwidth, and large dynamic range.

In the early stages, older VNAs, such as HP8753 or HP8510, prefer to adopt electrical samplers to realize the signals receiv-ing. The solution is phased out as the conversion efficiency degradation near the top of the sampler bandwidth limits the measurable frequency range [22], [34], [35]. Instead of the sampling solution, modern VNAs are mostly based on mixing solutions [21]–[23]. The mixer, as a basic component in the solutions, serves as downconverting high-frequency signals to intermediate frequencies for further processing and analysis. In practical applications, mixers need driving with a high power local oscillator power. For wideband and low-noise downconverting, the local oscillator should also have abilities in running over full frequency ranges and stability. It is challenging to realize all the abilities in the three aspects [23]. Thus, the harmonic mixing solution is proposed to limit the required frequency range of local oscillators, where high-frequency signals are mixed with harmonics of the local oscil-lator. However, new challenges come after, including conver- sion efficiency, local oscillator dependence, linearity, etc. [23], [36]. Moreover, parasitic capacitance, electromagnetic leakage, etc. in mixers lead to signal feedthrough between ports, which desensitize the receivers and compress the dynamic range [22], [37]. The rapid advance of microwave photonics technologies provides supportive and constructive solutions for the generation, processing, control, and distribution of microwave signals with wide bandwidth and high resolution [38]–[41]. A growing number of methods based on microwave photonics are proposed to overcome the challenges in the development of VNAs. M. Y. Frankel, et al. presented a VNA based on an impulse response record method in the time domain, where the impulse responses of the OUTs are recorded with the equivalent time sampling. Subsequently, the frequency domain amplitude-phase response is obtained by applying the Fourier transform on the impulse responses [42]–[44]. Similar to this idea, C. K. Lonappan, et al. improved the recording method with dispersion time-stretching scheme of chirped pulses, achieving a high rate sampling of the impulse responses [45], [46]. However, both of these two methods can not cover the tests of narrowband OUTs with a long-lasting impulse response. In that case, the former can introduce overlap interference between adjacent impulse responses [44]; the latter can truncate the impulse response since it exceeds the temporal width of the chirped pulse. The incomplete acquired impulse response can not provide complete frequency information. Besides, the two methods suffer from a low signal to noise ratio issue. The former has low modulation efficiency, while the usage of amplifiers in the structure further deteriorates the signal to noise ratio [44], [47]. The latter employs high multiple times stretching to slow down signals before digitization, inevitably resulting in power attenuation due to energy conservation [48]. Addi-tionally to impulse response measurement, A. R. Criado et. al. introduced a concept of heterodyne interface extension on commercial VNAs, without experimental verification [49]. In the concept, transmitted signals from devices under test are downconverted to an intermediate frequency by mixing with a selected harmonic frequency in an optical frequency comb and then be analyzed by the commercial VNAs. According to the author, the low conversion efficiency and the resulting low power values are two of the unsolved challenges. Sascha Preu presented a homodyne mixing concept [50]. Nevertheless, the proposed homodyne network analyzer, applying only one demodulation channel, is incapable of reading out both the amplitude and phase information from one DC term [51]. In the practical application, the homodyne detection, entailing IQ imbalance issues, is still not suitable for wideband VNAs [23].

In this article, we propose a photonic vector network analyzer (PVNA) with wideband photonic digitizing, which skips the down conversion process in the traditional VNAs and digitizes the reference and response signals directly. In the method, the OUT is stimulated by swept single tone signals. The reference signal, and the response signals, i.e. the reflected signal and transmitted signal, are sampled by an optical pulse train directly, followed by electro-optic conver-sion with photodiodes (PDs) and quantization with analog-to-digital converters (ADCs). The desired information about

OUTs, such as scattering parameters or other parameters based on it, is extracted from the digital sampling results with digital signal processing.

The proposed PVNA provides the following benefits. The stimulation with continuous waves enables the PVNA to dispense with the wide instant bandwidth, which allows the undersampling digitizing for the reference and response sig-nals with a low repetition rate of optical sampling pulse trains. Correspondingly, the OID, including PDs and ADCs, only acquire a bandwidth of half of the repetition rate, significantly reducing the necessity of the adoption of wideband devices. The replacement of the downconverting process with mixers with direct photonic digitizing excludes the challenges in the mixing solution and simplifies the system structure. On the other hand, the measurable frequency span of the method, or operational bandwidth, is determined by the temporal width of optical pulses and modulator bandwidth, able to extend to 110 GHz or higher with current commercial-off-the-shelf devices. The sweeping continuous-wave stimulation also makes it available to measure OUTs of any bandwidth without prior knowledge. Besides, the superior stability of the optical train ensures high sampling and measurement accuracy, thus avoiding the noise caused by sampling clock jitter. The adopted narrowband PDs and ADCs can provide a significantly low noise floor as well. On this basis, increasing the digitizing length can further improve the dynamic range to a high level.

## II. RESULTS

Theory and implementation. Figure 1 illustrates the conceptual diagram of the PVNA. The basic idea of the proposed method is to digitize the reference and response signals directly and analyze the characteristics of OUTs in the digital domain. After stimulated by a continuous wave, the OUT produces a corresponding response signal representing the inherent information. An ultrastable optical pulse train with a low repetition rate from a mode-locked laser (MLL) under-samples the response signal via an electrical-optical modulator (EOM). The femtosecond-level temporal width of the with optical pulses and ultrawideband electrical-optical modulator allow the measurable frequency range to extend to 100 GHz. An optical intensity digitizer (OID) translates the optical pulse train carrying the response signal into a digital signal, from which the digital signal processing can determine the inherent information at the given frequency. By sweeping the frequency of the incident signals, the responses in the specified frequency band can be obtained.

Figure 2 shows the experimental setup of the proposed PVNA, which involves four parts: a source module, a test set module, a receiver module, and an analyzer module. The source module generates a single-tone microwave signal to stimulate OUTs. The parameters of the single-tone signal(i.e. power, sweep range, the number of points, etc.) are specified by users. During the measurement, the frequency of the single-tone signal varies according to the specified sweep frequency range and points and the power keeps unchanged. The test set module routes the output from the source module to the OUTs and the receiver module further. A single-pole double-throw switch (SPDTS) is used to toggle between the two stimulated ports. Two power splitters (PSs) divide the signal from the source module into two parts: one part is an incident signal and stimulates the OUT; the other serves as a reference signal to assist in measuring the magnitude and phase of the incident signal. Two directional couplers (DCs) separate transmitted signals and reflected signals of the OUTs. When we connect the common pin and the T1 pin of the SPDTS, the incident signal stimulates the OUT from the port 1. Consequently, the reflected and transmitted signals are output from port 1 and port 2, respectively. Similarly, connecting the common pin and the T2 pin allows the stimulation of the OUT from port 2. The receiver module comprises four branches and measures the output signals from the test set module. Branch 1 and Branch 4 are reference branches that measure the reference signal. Branch 2 and Branch 3 are measurement branches that measure the reflected and transmitted signals. A mode-locked laser (MLL) generates an optical pulse train which is then sep-arated and fed into the four branches by a 1 4 optical coupler (OC). In each branch, an electrical-optical modulator (EOM), an optical intensity digitizer (OID) are cascaded in turn, where the OID consists of a photodiode (PD) and an analog-to-digital converter (ADC). The EOM acts as a sampling gate, allowing the optical pulse train to undersample the output signals from the test set module. Since the signals to be sampled is single-tone and have ultranarrow instant bandwidth, the signals are still be reconstructable without distortion at a low sampling rate according to Nyquist Shannon sampling theorem, which implies a low repetition rate optical pulse train can achieve ultra-high frequency measurements. The OID translates the intensity of each optical pulse into the digital data with a PD and an ADC. In detail, the PD converts the optical pulse train into an electric one and the ADC quantizes the amplitude of every electric pulse subsequently. The ADCs and the MLL are synchronized by a phase locked loop (PLL) so that each electrical pulse is quantized at the same location. The functions of the analyzer module include calculation and calibration. With digital signal processing (DSP) on the acquired digital data, the analyzer module completes the desired information calculation, such as scattering parameters. The calibration helps in mitigating the systematic errors and increasing the accuracy of measuring results. If the frequency of the incident signal is integer multiples of the MLL repetition rate, the sampling results are direct current and does not contain the valid information of magnitude or phase. To avoid this case, at the corresponding frequency, the analyzer module can adjust the MLL repetition rate by driving the piezoelectric transducer (PZT) in the MLL [52].

In the measurement branches of the test set module, after interacting with the OUT, a part of the incident signal is reflected while the rest is transmitted. The response signals, i.e. the reflected and transmitted signals, can be expressed as

$$v^0(t) = v(t) \; s_{ij}(t) \quad i = 1; 2 \; j = 1; 2 \; ; \qquad (1)$$

where $v(t)$ is the incident signal, and $s_{ij}(t)$ is the time domain response of the OUT to the incident signal. The i and j subscripts stand for the output and input ports, respectively. The asterisk operator denotes the convolution operation. In the reference branches of the test set module, the reference signal does not transmit through the device. In this case, the input signals to the receiver module are

$$v^0(t) = v(t). \tag{2}$$

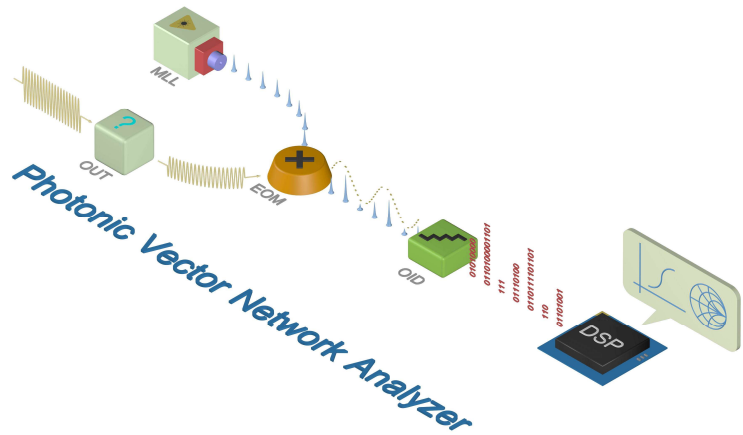

Fig. 1. Principle of the vector network analysis based on wideband direct photonic digitizing. MLL, mode-locked laser; OUT, object under test; EOM, electrical-optical modulator; OID: optical intensity digitizer; DSP, digital signal processing. A low repetition rate optical pulse train from a MLL undersamples the OUT's response signal via a wideband EOM. The OUT's response signal is extracted by the OID and determined with DSP. The OUT's response in the specified frequency band can be obtained by sweeping the frequency of the incident signals

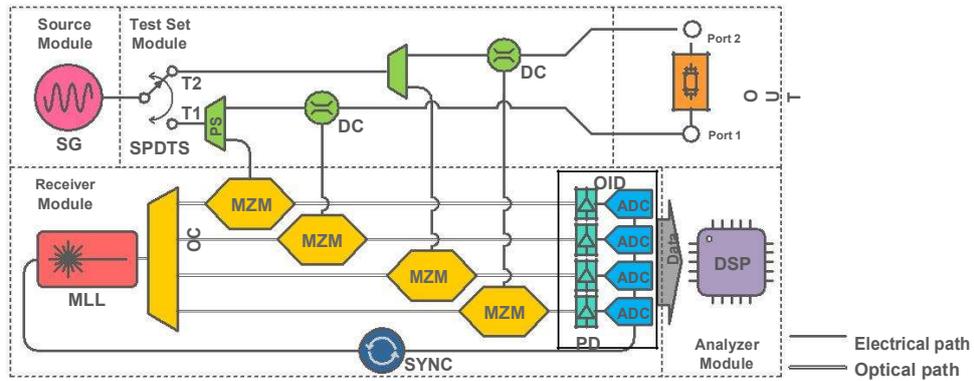

Fig. 2. Experimental setup of the proposed PVNA. SG: signal generator; SPDTS, single-pole double-throw switch; PS, power splitter; DC, directional coupler; OUT, object under test; MLL, mode-locked laser; OC, optical coupler; EOM, electrical-optical modulator; PD, photodiode; ADC, analog-to-digital converter; OID, optical intensity digitizer; SYNC, synchronization; DSP, digital signal processor

In the receiver module, each branch performs as a sampler, where the reference signal, reflected and transmitted signal from the test set module are sampled by the sampling optical pulse train and digitized with OIDs. The temporal shape of the sampling optical pulse train in a branch is

$$p(t) = P_A \sum_{k=1}^{\infty} p_S(t - kT_s), \tag{3}$$

where $P_A$ is the average pulse power of the sampling optical pulse train, $p_S(t)$ is the temporal power shape of a single optical pulse normalized by $P_A$. $T_s$ is the period of the sampling optical pulse train.

Within the specified compression power range, the sampling process can be viewed as a linear one [22], [23]. The amplitude of electric pulses output from PDs is proportional to the intensity of optical pulses in the case that the optical power is far less than the saturation power of PDs [53]. Therefore, the sampling results of a branch is

$$v_Q[k] = 0.5 \{[1 + h_M(t) * v^0(t)] p(t)\} * h_E(t - d_E)|_{t=kT_s}, \tag{4}$$

where $h_M(t)$ is the small-signal impulse response of the EOM, $h_E(t)$ is the impulse response of the OID including the PD and the ADC, and $d_E$ is a delay from the EOM to the ADC. As Eq. (4) indicates, a too high sampling rate and/or a too narrow OID bandwidth can lead to overlap between adjacent electrical pulses, interfering with the quantization on the electrical pulses and resulting in nonlinear distortion. The condition without interference is that the OID bandwidth should be greater than or equals to half of the sampling rate [54], [55].

After a derivation, we can see that the sampling results, $v_Q[k]$, involves two parts: an unmodulated component, $v_{Q0}[k]$, and a modulated component, $v_{Q1}[k]$ [54], [56], [57]:

$$\begin{cases} v_Q[k] = v_{Q0}[k] + v_{Q1}[k] \\ v_{Q0}[k] = 0.5 h_E(t - d_E) * p(t)|_{t=kT_s}, \\ v_{Q1}[k] = h_E(t) * h_M(t) * v^0(t) \end{cases} \tag{5}$$

where $h_A(t)$ is the equivalent channel impulse response whose expression is

$$h_A(t) = 0.5[h_E(t-d_E)p(-t)] * h_M(t). \quad (6)$$

The unmodulated component, $v_{Q0}[k]$, is independent of the signals to be sampled, $v^0(t)$, appearing as a direct current offset. Thus, subtracting the average from the sampling results can remove the unmodulated component. The modulated component, $v_{Q1}[k]$, contains the desired information of the OUT, $s_{ij}(t)$. From the equation the modulated component, we can consider each branch as an equivalent sampling channel with an equivalent impulse response, $h_A(t)$. In the equivalent sampling channel, the reference or response signal passes through the filter and sampler sequentially and becomes a digital copy. Figure 3 shows the equivalent sampling procedure, according to Eq. (5).

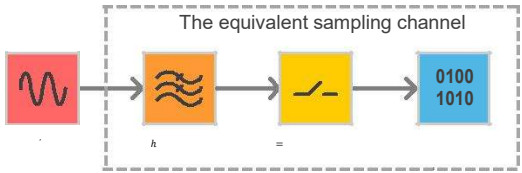

Fig. 3. Equivalent sampling procedure. The response signals are fed into an equivalent sampling channel whose impulse response and sampling period are $h_A(t)$ and $T_S$, respectively

In the sampling procedure, the output signal from $h_A(t)$ is a single-tone signal with a very narrow instant bandwidth, so that a low sampling rate is capable enough to obtain the complete magnitude and phase. When the frequency of the incident signal is within the first Nyquist zone of the sampler, the signal is sampled directly. However, when the frequency of the incident signal is beyond the first Nyquist zone of the sampler, the signal will be aliased down to the first Nyquist zone. In this process, the magnitude and the phase shift information can still be obtained from the aliased signal. Concretely, the output signal from $h_A(t)$ can be expressed in the frequency domain:

$$\begin{aligned} V_C(\omega) &= V^0(\omega)H_A(\omega) \\ &= A[\delta(\omega+\omega_0)\exp(-j\varphi) + \delta(\omega-\omega_0)\exp(j\varphi)], \end{aligned} \quad (7)$$

where $A$, $\omega_0$, and $\varphi$ are the magnitude, the analog angular frequency, and the phase of $v_C(t) = v^0(t)*h_A(t)$, respectively. After sampling, the signal can be represented in the digital frequency domain as

$$V_D(\omega) = \frac{1}{T_S}\sum_{l=-\infty}^{\infty} V_C\left(\frac{\omega}{T_S} - \frac{2\pi l}{T_S}\right), \quad (8)$$

where $\omega = \Omega T_S$ is the digital angular frequency. The range of the digital spectrum is from $-\pi$ to $\pi$. For the signal with a frequency of $\omega_0$ within the range of $(m\omega_s; (2m + 1)\omega_s/2]$ and a phase of $\varphi$, the digital spectrum of it in this observation range is

$$V_D(\omega) = A[\delta(\omega+\omega_0)\exp(-j\varphi) + \delta(\omega-\omega_0)\exp(j\varphi)] \quad (9)$$

where $m$ is an integer and $\omega_0 = \Omega_0 T_S - 2\pi m$ is the digital frequency after undersampling. From Eq. (9), we can tell that the aliasing does not change the magnitude and the phase of the sampled signals. On the other hand, for the signal of $\Omega_0$ within frequency within $((2m + 1)\omega_s/2; (m + 1)\omega_s]$ and a phase of $\varphi$, the digital spectrum of in the observation range is

$$V_D(\omega) = A[\delta(\omega+\omega_0)\exp(j\varphi) + \delta(\omega-\omega_0)\exp(-j\varphi)]. \quad (10)$$

where $\omega_0 = 2\pi m - \Omega_0 T_S$ is the digital frequency after undersampling. Equation (10) tells that the aliasing does not change the magnitude of the sampled signal, however, reverse the phase. Therefore, we can obtain the magnitude and the phase of the signal with DPS techniques, and correct the phase with phase reversal at the corresponding frequencies.

Besides the OUT characteristics, the measurement results obtained in the measurement branch also contain the information on the incident signal which should be excluded. To exclude the magnitude of the incident signal, we can calculate the ratio of the magnitude of the reflected wave or transmitted signal from the measurement branch to that of the incident signal from the reference branch. As for the phase shift, in the measurement branch, the phase calculation result after sampling includes three parts: the initial phase of the single-tone signal, the phase shift introduced by the measurement branch, and the phase shift caused by the OUT. Since the initial phase of the single-tone signal generated by the source module is uncontrollable, the other reference branch is needed to assist in determining the phase shift caused by the OUT. In the reference branch, the sampled signal is generated simultaneously with the signal in the measurement branch, but independent of the OUT, thus the phase calculation results of which only include two parts: the initial phase of the single-tone signal and the phase shift introduced by the phase reference branch. The influence of the initial phase can be eliminated by comparing the phase calculation results in the measurement branch with the phase calculation results in the reference branch. The difference between the phase shift introduced by the measurement branch and the reference branch at the given frequency is a constant, which can be removed by calibration.

Calibration is an indispensable process to remove systematic errors. The systematic errors are caused by imperfections in the test setup and contribute to repeatable and predictable measurement errors. As Fig. 4 shows, the imperfections include directivity errors caused by directional couplers, crosstalk relating to signal leakage, source, and load impedance mis-matches relating to interactions between the test system and the OUT input/output match, frequency response errors caused by reflection and transmission tracking within branches in the receiver module. The imperfections cause systematic errors in PVNA are the same as ones in traditional VNAs [21], [58], [59], implying the measuring results can be calibrated by several proposed calibration methods in traditional VNAs. Operational bandwidth. In order to sweep OUTs in a large frequency span, the operational bandwidth of the equivalent sampling channel should be sufficiently wide. The equivalent frequency response of the sampler can be obtained from Eq.

(6) with Fourier transform of $h_A(t)$:

$$H_A(\omega) = 0.5P_A H_M(\omega)P_S(\omega)R(\omega), \quad (11)$$

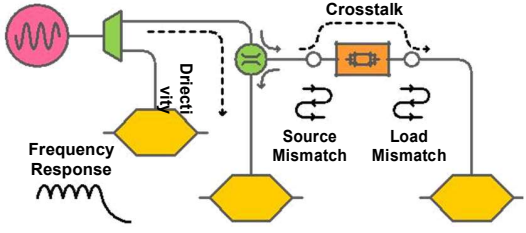

Fig. 4. Schematic of systematic measurement errors in PVNAs. The sys-tematic measurement errors include directivity error, crosstalk, source mis-match, load mismatch, reflection frequency response error, and transmission frequency response error

where $H_M(\omega)$ is the small-signal frequency response of EOM with cables, $P_S(\omega)$ is the Fourier transform of temporal power shape of a single optical pulse, and $R(\omega)$ denotes the effect of inter electrical pulse interference in the channel. $R(\omega)$ is expressed as

$$R(\omega) = \frac{1}{T_s} \sum_{n=1}^{\infty} \times H_E(\omega + n\omega_s)\exp[-j(\omega + n\omega_s)\Delta t_E]; \quad (12)$$

where $H_E(\omega)$ is the Fourier transform of $h_E(t)$. If the inter electrical pulse interference does not occur, $R(\omega)$ is a constant, and the channel frequency response can form a continuous passband. Nevertheless, $R(\omega)$ is a periodic function with a period $\omega_s$, leading to periodic ripples on the system frequency response and a limitation on the system bandwidth. Therefore, for a PVNA without the inter electrical pulse interference, the OID bandwidth will not limit the system bandwidth, which indicates that narrowband PDs and ADCs can also achieve an ultra-wide measurement bandwidth in a PVNA. Additionally, in the case without interference, the measurement bandwidth is determined by the product of $H_M(\omega)$ and $P_S(\omega)$. At present, a number of MLLs generating a temporal width of less than 100 fs have been studied well and entered the commercial stage [60]–[65]. Meanwhile, EOMs, including commercial ones, have also reached a bandwidth beyond 100 GHz [66]–[69]. The development of these optical devices enables the proposed PVNA to measure ultra high frequency characteristics of OUTs.

Dynamic range. Dynamic range is an essential parameter for measurements on high-dynamic-range components. Equations (5) and (6) indicate, on the one hand, the average optical pulse power, $P_A$, and the responsivity of the OID, $h_E(t)$, determine the maximum magnitude of the quantized digital signal. On the other hand, the linear operating range of the EOM limits the amplitude of the input microwave signal. Thus, the product of these three parameters determines the upper limit of the dynamic range. Moreover, the noise source (mainly the PDs in OID) can generate noise and pollute the sampling results, affecting the information extraction. After generated, the noise from PDs is added to the electric pulses and then quantized together into digital signals, thus treated as additive [70]. The power level of the noise floor is considered as the lower limit of the dynamic range since a signal with power less than the noise floor will be buried in noise. Therefore, we can expand the dynamic range by improving the upper limit and reducing the lower limit of the dynamic range. Correspondingly, increasing the average pulse optical power and responsivity of the OID and linearization of EOM are three alternative options. In terms of noise floor reduction, besides using devices with low noise power density, we also can increase the resolution bandwidth by increasing the number of Fast Fourier Transform (FFT) points. When a digital signal is transmitted into the frequency domain with an N-point FFT, it can be considered as sent through a bank of N filters with a bandwidth or frequency bin spacing of $\Delta f = f_s/N$. The filters can have a narrower bandwidth and the power at each bin becomes smaller as N or the number of frequency bins is increased. Since the sampled signal is a single tone single, the power of the sampled signal is irrelevant to the number of frequency bins. However, the noise has wide bandwidth and the noise floor after FFT is can be suppressed by increasing the number of frequency bins. Therefore, in this case, the dynamic range can be improved at a rate of $10\log_{10}(N=2)$ in the decibel unit [71].

Experiment. To investigate the performance of the proposed PVNA, an experiment is performed. The incident signal is generated from a source module (Rohde & Schwarz, SMF 100A). The bandwidth of the power splitters and used direc-tional couplers in the test set module are all beyond 40 GHz. An MLL (Precision Photonics, FFL1560) generates optical pulses at a repetition rate of 36.456 MHz. A 1:4 optical coupler splits optical pulses into four branches and then routes them into 40 Gbps quadrature-biased Mach-Zehnder modula-tors respectively. After that, optical pulses are converted into electric pulsed by PDs and quantized by an ADC (Keysight, M9703A). The bandwidths of PDs and ADC are 300 MHz and 1.2 GHz, respectively. Since the bandwidth of PDs is much lower compared with the bandwidth of the ADC, the overall OID bandwidth is limited to 300 MHz. During the measurement, we use a bandpass filter with a center frequency of 35 GHz as the OUT. The measuring frequency range is set from 30 GHz to 40 GHz.

The optical pulses generated from the MLL have a temporal width of 500 fs, corresponding to a bandwidth of 600 GHz, which is measured with a frequency-resolved optical gating (Swamp Optics, FROG). The frequency response of the used MZM with the cables and connectors is also shown in Fig. 5, which indicates the bandwidth is 20 GHz. The bandwidth of the MZM is much narrower than that of the optical pulses, so that the operational bandwidth of the established PVNA is limited by the bandwidth of the MZM in theory. Figure 5 shows the system frequency response of the established PVNA, which is measured by connecting port 1 and port 2 directly without calibration. It can be seen that the system frequency response is restricted by the the frequency response of the the used MZM with the cables and connectors, as the theoretical analysis. Moreover, since the overall OID bandwidth is more than 8 times that of the repetition rate, the requirement in the non-interference condition can be satisfied [54], [55]. Therefore, the system frequency response forms a continue passband without ripples. The maximum attenuation of the established PVNA in the range of 0 to 40 GHz is 7.5 dB.

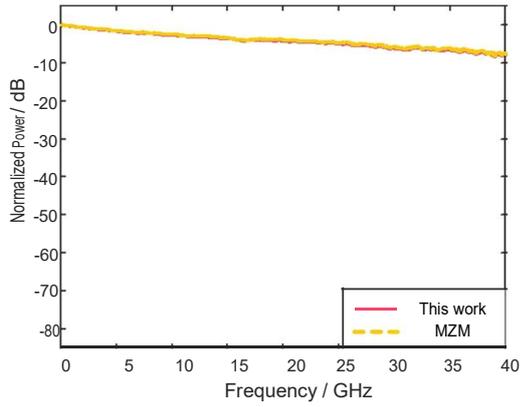

Fig. 5. The system frequency response of the established PVNA (solid line) and the used MZM with the cables and connectors (dashed line). Since the bandwidth of optical pulse is much larger than that of the MZM and the OID bandwidth satisfies the non-interference condition, the operational bandwidth of the established PVNA depends on the bandwidth of the MZM

Figure 6 shows the measuring results of phase shift between the measurement branch and the reference branch, where the measuring frequency range from 34.5 GHz to 35 GHz and the incident power is set to 0 dBm to ensure the system linearity. Since the frequency of the sampled signal is far more than half of the sampling rate, the undersampled signal is aliased to the first Nyquist zone. The aliasing causes the phase reversal at the period of 18.228 MHz, i.e. a half of the sampling rate, as Eqs. (9) and (10) indicate. After reversing the phase within frequency within $((2m+1)\omega_s/2; (m+1)\omega_s]$, the phase shift is corrected to a linear one, shown as a yellow line in Fig. 6.

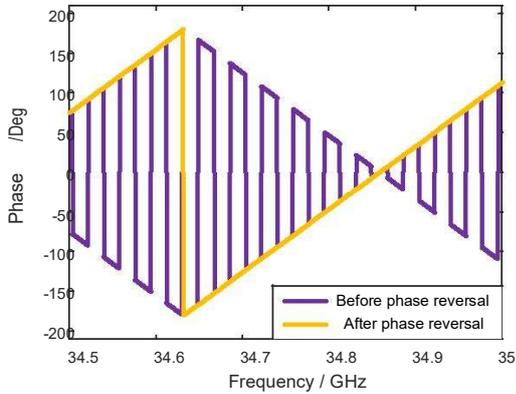

Fig. 6. Phase shift measuring results before phase reversal (purple line) and after phase reversal (yellow line). The undersampling introduces phase aliasing and reverses the phase shift at the period of 18.228 MHz, a half of the sampling rate. Reversing the phase within frequency within $((2m+1)\omega_s/2; (m+1)\omega_s]$ correct the phase shift measuring results into a linear one

In vector network analysis, the system linearity is described by 0.1-dB compression point and one of the limits on dynamic range. In the experiment, MZM is chose as the EOM, whose linearity is limited by its nonlinear transmission function. Assuming that the sampled signal has a cosine form, by applying Jacobi-Anger expansion, the output of the modulator biased at quadrature point can be expressed with Jacobi-Anger expansion as:

$$\begin{aligned}p_m(t) &= \tfrac{1}{2}p(t)\left[1+\cos\left(\tfrac{\pi A}{V_\pi}\cos(\omega t)+\tfrac{\pi}{2}\right)\right]\\&=\tfrac{1}{2}p(t)\{1-\sin[\tfrac{\pi A}{V_\pi}\cos(\omega t)]\}\\&=\tfrac{1}{2}p(t)-p(t)\sum_{n=1}(-1)^n J_{2n-1}(\tfrac{\pi A}{V_\pi})\cos[(2n-1)\omega t]\end{aligned} \quad (13)$$

where $V_\pi$ is the MZM half-wave voltage and $J_{2n-1}(\cdot)$ is the $2n-1$-th Bessel function. Equation (13) implies that the 0.1-dB compression point of the fundamental is determined by the half-wave voltage. The half-wave voltage of the adopted MZM is $\sim 5.4$ V in the experiment, corresponding to a 0.1-dB compression point of $\sim 5.6$ dBm in the case of 50 impendence match. To demonstrate the linearity performance of the proposed PVNA, we implement a power sweep measurement with a 35 GHz signal. As Fig. 7 illustrates, the 0.1-dB compression point in the established PVNA is 2.8 dBm. The difference from the theoretical analysis can be resulted by nonlinear responsivity of PDs.

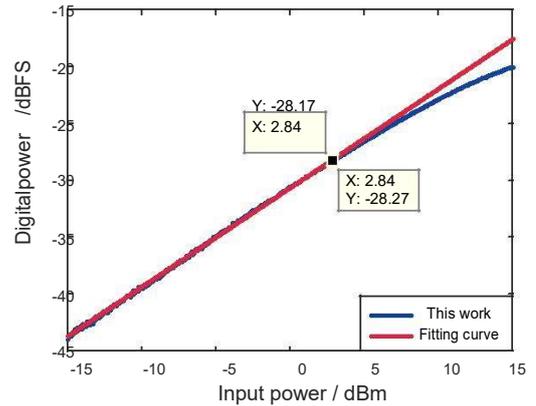

Fig. 7. 0.1-dB compression point of the established PVNA is 2.8 dBm.

Figure 8 shows the measuring digital power spectrum of a 35 GHz single tone singnal with different numbers of FFT points, where the frequency is normalized by the sampling rate. The 35 GHz single tone singnal is aliased to a normalized frequency of 0.07. In the figure, when the number of FFT points is 6.25e4, the level of noise floor is 102 dB. In the case of increasing the number of FFT points from 6.25e4 to 2.5e5, 1e6 and 4e6, the level of noise floor corresponds to 108 dB, 114 dB and 120 dB, respectively. It can be seen that for every four times increase in the number of FFT points, the noise floor is reduced by 6 dB and the dynamic range is increased by 6 dB, correspondingly. During the increase of the number of FFT points, the power of the aliased signal keeps unchanged. The proposed PVNA achieves a dynamic range of 120 dB under the condition of 35 GHz signal excitation at 0.1-dB compression point and 4e6-point FFT.

The measured scattering parameters results of PVNA and the commercial VNA (Keysight, PNA-X N5247A), including both the magnitude and the phase shift, calibrated with a Short-Open-Load-Through (SOLT) calibration via a standard

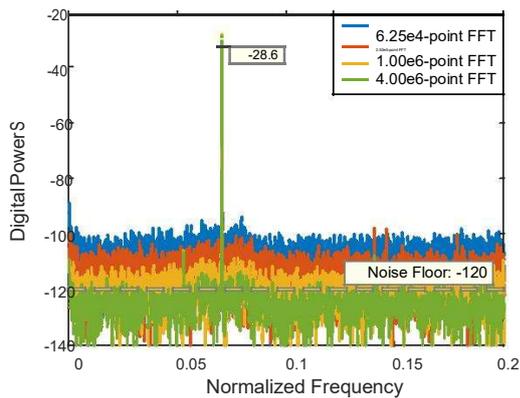

Fig. 8. Dynamic range measurement of the established PVNA with 6.25e4, 2.5e5, 1e6 and 4e6 points FFT. Four times increase in the number of FFT points can extend the dynamic range by 6 dB. A 120 dB dynamic range is achieved under the condition of 35 GHz 2.8 dBm signal excitation and 4e6-point FFT

calibration kit (Keysight 85056D) [58] are illustrated in Figure 9. From the obtained $S_{12}$ or $S_{21}$, one can find that the center frequency of the OUT is 34.725 GHz and the 3 dB bandwidth is 4.25 GHz. At the center frequency, the voltage standing wave ratio (VSWR) is 1.5 from the obtained $S_{11}$ or $S_{22}$. By deriving the phase shift versus frequency, the delay of the OUT for different frequencies can be obtained. From the obtained $S_{21}$ phase shift, the average delay of the OUT within the passband is calculated as 900 picoseconds.

## III. DISCUSSION

We proposed a vector network analyzer based on wideband direct photonic digitizing. Utilizing the undersampling with ultrashort optical pulses, electro-optic conversion with photodiodes and quantization with ADCs, the proposed photonic vector network analyzer realizes a direct digitizing of the reference and response signals of OUTs. The proposed PVNA not only enables the measurement with high accuracy, linearity and dynamic range but also discards the complicated downconversion procedure in conventional electrical VNAs. Furthermore, we established PVNA covering a measurable frequency span of up to 40 GHz and a dynamic range of 120 dB and demonstrated scattering parameters measurement of a bandpass filter experimentally. The proposed method has the potential to be a prevailing one for research, development, evaluation, and testing of a variety of fields such as medicine, material, geology, communication and so on.

## ACKNOWLEDGEMENTS

This work was supported in part by the National Natural Science Foundation of China (NSFC) (61535006, 61627817).

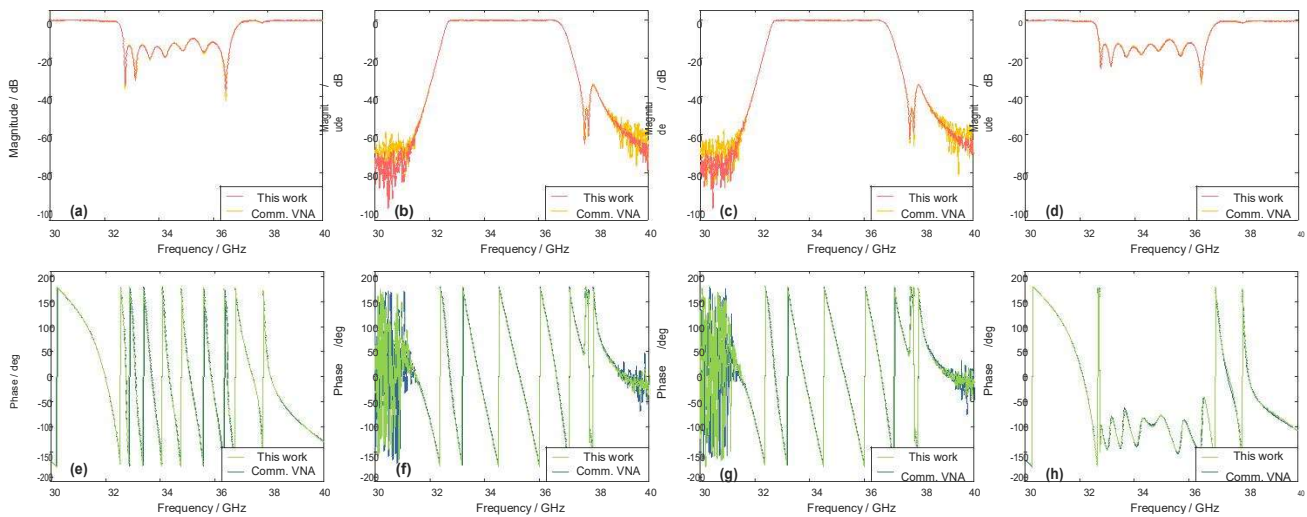

Fig. 9. Measuring results. (a)-(d): |S11|, |S12|, |S21|, and |S22|; (e)-(h): ∠S11, ∠S12, ∠S21, and ∠S22